%% file: paper-loyalty.tex
\def\year{2017}\relax

\documentclass[letterpaper]{article}
\usepackage{aaai17}
\usepackage{url}
\usepackage{graphicx}
\usepackage[utf8]{inputenc}
\usepackage{amsmath}
\usepackage{amsfonts}
\usepackage[T1]{fontenc}
\usepackage{txfonts}
\usepackage{color}
\usepackage{booktabs}
\usepackage{enumitem}

\usepackage{amssymb}

\newcommand*\samethanks[1][\value{footnote}]{\footnotemark[#1]}
\newcommand{\xhdr}[1]{{\noindent\bfseries #1.}}
\newcommand{\sub}{}
\newcommand{\cut}[1]{}
\newfont{\eaddfnt}{phvr8t at 9pt}
\author{
{\normalsize
William L. Hamilton{\Large\bf \thanks{The two first authors made equal contributions and are ordered alphabetically.}}$^\dagger$, Justine Zhang{\Large\bf  \samethanks[1]}$^\ddagger$, Cristian Danescu-Niculescu-Mizil$^\ddagger$, Dan Jurafsky$^\dagger$, Jure Leskovec$^\dagger$
}\\
$^\dagger$Stanford University, $^\ddagger$Cornell University\\
{\eaddfnt{wleif@stanford.edu, jz727@cornell.edu, cristian@cs.cornell.edu, jurafsky@stanford.edu, jure@cs.stanford.edu}}
}
\title{Loyalty in Online Communities}
\begin{document}

\maketitle
\input{000abstract}
\input{010intro}
\input{020relatedwork}
\input{030measuring_loyalty_REVISED}
\input{050community}

\input{040individual}
\input{070conclusion}
\section*{Acknowledgements} 
The authors thank Lillian Lee, Tianze Shi, Skyler Seto, David Jurgens, and Vinod Prabhakaran for their helpful comments.
This research has been supported in part by a Discovery and Innovation Research Seed Award from the Office of the Vice Provost for Research at Cornell, NSF
CNS-1010921,     
IIS-1149837, IIS-1514268
NIH BD2K,
ARO MURI, DARPA XDATA,
DARPA SIMPLEX, DARPA NGS2,
Stanford Data Science Initiative,
SAP Stanford Graduate Fellowship, NSERC PGS-D,
Boeing,          
Lightspeed,			       
and Volkswagen.  
\bibliographystyle{aaai}
\bibliography{loyalty}
\end{document}

%% file: 000abstract.tex

\begin{abstract}
Loyalty is an essential component
of multi-community engagement. 
When users have the choice to engage with a variety of different communities, they often become loyal to just one, focusing on that community at the expense of others.
However, it is unclear how loyalty is manifested in user behavior, or whether certain community characteristics encourage loyalty.

In this paper we operationalize loyalty as a user-community relation: \emph{users loyal} to a community consistently prefer it over all others; \emph{loyal  communities} retain their loyal users over time.  
By exploring a large set of Reddit communities,
we reveal that loyalty is manifested in remarkably consistent behaviors. 
Loyal {\em users} employ language that signals collective identity and engage with more esoteric, less popular content,
indicating that they may play a curational role in surfacing new material. 
Loyal {\em communities} have denser user-user interaction networks and lower rates of triadic closure, suggesting that community-level loyalty is associated with more cohesive interactions and less fragmentation into subgroups. 
 We exploit these general patterns to predict future rates of loyalty.
 Our results show that a user's propensity to become loyal is apparent from their initial interactions with a community, suggesting that some users are intrinsically loyal from the very beginning. 
\end{abstract}

%% file: 010intro.tex

\section{Introduction}
The Internet offers a staggering variety of virtual communities for the intrepid wanderer to explore.
Faced with this abundance of options, a user may have fleeting relationships with some communities, choosing to allocate only a small proportion of her time to each \cite{tan_all_2015}. 
Alternatively, a user may commit to forming a more steadfast relationship with one particular community, establishing her loyalty to that group by consistently preferring it above all others. 

While there is a rich literature studying various flavors of user-to-user relationships within individual communities \cite{mckenna_causes_1999,fiore_observed_2002,arguello_talk_2006,danescu-niculescu-mizil_mark_2011,ellis_equality_2016}, 
less focus has been directed at understanding relationships that exist between {\em users} and {\em communities}.

Loyalty is a fundamental example of such a relationship.
 In multi-community platforms like Reddit, users have no shortage of alternative communities to peruse \cite{tan_all_2015,hessel_science_2016}, so understanding why a user chooses to be loyal to one community and not to others is a central problem. 
Characterizing these loyal relationships in terms of the traits of their two components---users and communities---can 
 offer insights into how community identity arises online \cite{cassell_language_2005,nguyen_language_2011}, and also guide community maintainers towards fostering better user involvement, with their most faithful users in mind.

Loyalty is fundamentally about the preferences and commitments of active, engaged users, and while there has been considerable effort spent studying how to attract and retain new users \cite{karnstedt_churn_2010,dror_churn_2012} or improve user engagement \cite{arguello_talk_2006,backstrom_characterizing_2013}, there is comparatively little understanding of how already-active users choose to allot their time between communities \cite{tan_all_2015}.
In particular, loyalty 
can
 only exist within the context of multi-community dynamics:
 unlike the well-studied issue of
 user churn or retention, which is usually defined in terms of a single, isolated community \cite{dror_churn_2012,danescu-niculescu-mizil_no_2013}, understanding loyalty requires examining user preferences in the context of multiple communities.

 \xhdr{Present work} 
Our aim is to provide a thorough characterization of user loyalty in the 
multi-community platform, Reddit.
To this end, we develop a measure of loyalty in terms of user preference and commitment; loyal users prefer one community above all others and maintain this preference for a substantial period of time. 
Applying this measure to Reddit, we characterize loyal {\em users}, as well as {\em communities} that foster user loyalty.   
The large-scale, multi-community nature of Reddit, where users can peruse thousands of 
communities, makes it an ideal setting to study 
loyal behavior.

We reveal behavioral traits that systematically distinguish loyal users across a diverse set of communities, and show that loyal communities share similar structural features.  Together these observations provide a cross-community characterization of loyalty that lends a new perspective on user-community dynamics.  

First, we characterize loyalty at the community level.
We show that topics that reflect strong external interests, like sports and videogames,  tend to engender high rates of loyalty in communities.
Moreover, we show that
beyond topical characterizations, there are structural, social network features that distinguish highly loyal communities. 
We find that loyal communities tend to have denser user-user interaction networks, even after controlling for community-wide activity levels.
The interaction networks of loyal communities are also less clustered and contain more bridging ties connecting active and inactive users. 
Overall, our results suggest that loyalty is conducive to, and thrives in, communities that are more inclusive and cohesive.

Second we find that, at the individual-level, loyal users exhibit many characteristic behavioral patterns that generalize across communities.
For example, loyal users tend to comment on less popular
 and more esoteric 
 posts, and do so using language that signals collective identity.  
These features are even exhibited in the very first contributions a user makes to a community, suggesting that certain users are already loyal from the very beginning.
The fact that (future) loyalty can be detected in users' first contributions also shows that the features we uncover are {\em predictive} of loyalty and not simply the result of users having repeated interactions with the community.
We exploit this observation in the task of predicting which users will eventually become loyal, a task of practical importance to community maintainers.

%% file: 020relatedwork.tex

\section{Related Work}

Research on loyalty and associated concepts spans a wide-variety of disciplines, including sociology \cite{connor_sociology_2007}, social psychology \cite{van_vugt_social_2004}, and marketing \cite{chaudhuri_chain_2001}.

\xhdr{Theories and studies of loyalty}
In social psychology, phenomena related to loyalty pertain to social identity theory and its counterparts \cite{tajfel_social_2010}, which address the interface between social identity and overt manifestations of loyal behavior \cite{zdaniuk_group_2001,van_vugt_social_2004}.

One area in which loyalty has received considerable direct attention is marketing, where the notion of brand loyalty plays a central role \cite{tucker_development_1964,jacoby_brand_1971,day_two-dimensional_1976,chaudhuri_chain_2001}.
Unlike work in the previously mentioned areas, marketing specialists have spent considerable effort on quantifying objective measures of loyalty in a data-driven manner (\citeauthor{jacoby_brand_1971} \citeyear{jacoby_brand_1971}; \citeauthor{day_two-dimensional_1976} \citeyear{day_two-dimensional_1976}; {\em inter alia}).
Brand loyalty has been measured in numerous ways, but almost all approaches somehow quantify the extent to a which customer repeatedly purchases one brand 
over its competitors; see Jacoby and Chestnut \shortcite{jacoby_brand_1978} and Odin et al \shortcite{odin_conceptual_2001}	for comprehensive surveys.
Our operationalization of loyalty draws inspiration from such characterizations.

\xhdr{Loyalty in online communities}
Within the study of online communities there has been little direct work on loyalty. 
Notable exceptions are Sharara et al  \shortcite{sharara_understanding_2011}, who develop a measure of loyalty for dynamic affiliation networks, and Newell et al \shortcite{newell_user_2016}, who investigate platform-level loyalty through a case-study of a large-scale user migration event.
We substantially extend these works by uncovering common, predictive traits that are indicative of both user- and community-level loyalty. 

There are numerous works exploring the related phenomenon of user retention or churn in online communities \cite{arguello_talk_2006,karnstedt_churn_2010,dror_churn_2012,ngonmang_churn_2012,oentaryo_collective_2012,danescu-niculescu-mizil_no_2013}.
However, predicting user churn is fundamentally distinct from the more nuanced concept of loyalty, which emphasizes the preferences and commitments of active users in a {\em multi-community} environment. 

The concept of loyalty can also be viewed as a specific form of user engagement \cite{lampe2010motivations,zhang2017typology}.
Unlike previous work on user engagement---which often focuses on characterizing and steering short-term engagement within a single community \cite{rashid_motivating_2006,anderson2013steering}---our analysis of loyalty focuses on longer-term participation and explicitly addresses user preferences within a large-scale, multi-community setting. 
In other words, instead of modeling {\em how much} a user engages with a single community, we model {\em the proportion} of activity that an individual user chooses to allocate to that community, above others. 

Our work is also closely related to studies of  socialization, acculturation, and sociolinguistics in online communities \cite{cassell_language_2005,danescu-niculescu-mizil_mark_2011,nguyen_language_2011,danescu-niculescu-mizil_no_2013}.
None of these works have directly studied loyalty per se, but a number of the user-community dynamics uncovered in these are relevant here.
In particular, these studies have revealed important stylistic changes in user writing---such as a decreased use of 
first person pronouns over time---as users become more ``socialized'' in a community. 
We uncover similar stylistic markers of loyalty in the present work. 

%% file: 030measuring_loyalty_REVISED.tex

\section{Operationalizing Loyalty}\label{sec:defining-loyalty}

We start by conceptualizing user-community loyalty at a high level, motivated by prior work in marketing and social psychology. We then describe how we operationalize a measure of loyalty in Reddit, the multi-community platform studied in this paper.

\subsection{Motivating a definition of loyalty} 
\label{sub:motivating_a_definition_of_loyalty}

We define loyalty as a combination of {\em preference} and {\em commitment}.
Loyal users exhibit a clear preference for one community and sustain this preference over time.

Various formal and intuitive definitions of loyalty exist \cite{fletcher_loyalty:_1995}.
However, for the task of quantifying loyalty in a multi-community setting, our approach---which combines both preference and commitment---has a number of benefits.
For example, we could simply define users as loyal if they make consistent contributions to a community over time; however, this approach is confounded by baseline differences in activity levels and would improperly assume that all highly active users are also loyal.
Another alternative would be to define users as loyal if they comment substantially more in one community compared to others, ignoring temporal considerations; however, such a time-agnostic approach could be easily confounded by transient fads. 
In our definition, we measure the relative extent to which users prefer one community over others, disentangling the notion of user-community loyalty from baseline rates of user engagement, and we require that this preference is sustained over time, ensuring that we measure actual commitment and not just transient interest. 

Lastly, while some notions of loyalty permit individuals to be loyal to multiple groups at the same time \cite{fletcher_loyalty:_1995}, we choose to study an exclusive variant where users can be loyal to at most one community at a point in time.
This choice has both theoretical and practical motivations.
From a theoretical perspective,
we want to know which community a user is most attached to (not a set of communities they interact with), since this primary community is likely to more strongly influence the user's online (Reddit) identity, as well as how they interact with other communities \cite{hewstone_intergroup_2002,tajfel_social_2010}.
From a practical perspective, allowing users to be loyal to multiple communities would also introduce unnecessary complications, requiring various activity-based thresholds to determine which communities are included in this ``loyal set''.  
Opting for an exclusive variant of loyalty simplifies our analyses by focusing them on 
the strongest user-community relationships. 
That said, definitions of loyalty that allow users to be loyal to multiple groups, along with other reasonable variations, could prove useful for certain applications, and exploring this space is an interesting direction for future work.

\subsection{Operationalizing loyalty on Reddit} 
\label{sub:defining_loyalty_on_reddit}


We perform our analysis on a dataset of posts and comments from Reddit, a popular website where users form topical discussion-based communities (called subreddits). 
Users can submit \emph{posts} to a subreddit, consisting of a post title along with urls, images, and/or explanatory text; users can also \emph{comment} on existing posts and reply to each others' comments in a thread-based interface. Posts and comments also receive scores, or votes, that serve as a form of community feedback. 

Reddit contains thousands of active communities that are constantly changing, with new user-defined communities arising daily \cite{tan_all_2015}. 
The large number of possible communities to explore makes Reddit an ideal dataset for studying loyalty.
A loyal Reddit user must decide to consistently allot time to a particular community, despite an ample availability of often closely related alternatives \cite{hessel_science_2016}.

Our full dataset consists of all comments and posts made to Reddit in 
2014: approximately $10^{8}$ comments made by $10^{7}$ users in $10^{4}$ communities.\footnote{\url{https://archive.org/details/2015_reddit_comments_corpus}. For computational simplicity, we discard the long-tail of inactive communities that have less than 250 users per month.} 
The following subsections describe the subsets we use in different parts of our analysis. 

\xhdr{Loyal users}
We define user loyalty on Reddit based on commenting behavior, which we view as a strong proxy for latent engagement. 
To focus our attention on loyalty instead of platform-level retention, all of the following definitions are restricted to users who commented at least 10 times within the relevant time period. 
Additionally, we only consider {\em top-level comments} that are initial responses to a post.
Top-level comments more clearly
demonstrate a user’s evaluative choice to comment in a particular community compared to lower-level comments, which may result from the social obligation to maintain a conversation.

We say that a user $X$ {\em prefers} a community $A$ in month $t$ if at least 50\% of the comments that $X$ submits across Reddit in $t$ are to $A$. $X$ is then {\em loyal} to $A$ at $t$ if $X$ prefers $A$ at both $t$ and $t+1$ (i.e., exhibits commitment). We use monthly time windows, following common practice in studies of user engagement and churn (\citeauthor{oentaryo_collective_2012} \citeyear{oentaryo_collective_2012}; \citeauthor{danescu-niculescu-mizil_no_2013} \citeyear{danescu-niculescu-mizil_no_2013}; {\em inter alia}).
Note that this definition is specific to a particular month, so a user is loyal at a particular point in time $t$ and her loyalty can shift over time. 
For simplicity, we use the phrase ``loyal users'' to refer to the set of users who were loyal to a community at some point in time, and when examining the behavior of a loyal user, we only use data from the time-period in which she was loyal. 

To provide a reference point for the behavior of loyal users, we contrast loyal users with {\em vagrant} users who fleetingly interact with a community before wandering off. 
We define a {\em vagrant} of $A$ as a user who comments between 1 and 3 times in $A$ at $t$, and, while still active on Reddit at time $t+1$, does not contribute to $A$ in $t+1$. 
This definition of vagrant users ensures that we are comparing loyal users to other active Reddit users who interacted with community $A$.

In order to have enough statistical power for within-community analyses, we restrict our user-level studies to communities with at least 25 loyal and 25 vagrant users per month. 
This results in $242$ communities covering a diverse range of populations and topics, from videogame communities like \sub{/r/starcraft} to religious communities like \sub{/r/Catholicism}.
In total our analysis set contains 177,593 loyal users and 1,989,530 vagrants, with the median community containing 353 and 3,046 of each category, respectively. 
Across all communities in each month, these loyal and vagrant users contribute about 10\% each of all of the comments made to a community.

\xhdr{Loyal communities}
 \newcommand{\lr}{loyalty-rate}
\newcommand{\unloyal}{non-loyal}
As a natural extension of our user-level definition, we say a {\em community} is loyal if it tends to retain a high proportion of loyal users month after month. 
We focus on success at retaining loyal users, rather than counts of loyal users at particular points in time, in order to minimize confounds due to differences in community sizes and popularities. 

We compute a {\em \lr} for each community $A$ as the expected proportion of users who prefer\footnote{To achieve sufficient statistical power, in community-level experiments we only require that users comment in $A$ more than any other community that month, without the 50\% threshold used in the user-level experiments.
 Analogous trends hold without this relaxation but do not reach the same significance levels.} $A$ at $t$ and sustain this preference at $t+1$. 
Users who prefer $A$ at $t$ and then leave Reddit altogether are ignored, since we seek to model inter-community loyalty and not platform-level churn.  
By focusing on month-to-month dynamics and ignoring longer time-scales, this definition makes a Markov-esque assumption regarding user behavior. 
Nonetheless, we generally find that in communities with higher loyalty rates, individual loyal users also tend to stay for more months in total, indicating that these monthly transition rates do signal longer-term commitment (Spearman's $\rho=0.53, p<10^{-10}$, comparing loyalty rates with the average tenure of loyal users). 

We analyze all communities with at least 25 loyal users in one month, resulting in 1440 communities (with a median loyalty rate of $60.7\%$).
For the purpose of this study, we denote \emph{loyal communities}, as the top-25\% of this distribution and \emph{\unloyal} communities as the bottom-25\%. 

%% file: 050community.tex

\section{Community-level Loyalty}\label{sec:communities}

We begin by analyzing the types of communities that tend to foster high rates of loyalty. 
We find that loyal communities exhibit consistent structural features in their user-user interaction networks, and that these structural features are predictive of loyalty,
across communities with vastly different topical interests.

\subsection{Basic features of loyal communities}
Loyal communities are significantly smaller than \unloyal\ communities ($p<10^{-5}$, U Test); the median loyal community is 39\% smaller than the median \unloyal\ one. 
For example, many communities that are highly successful at retaining loyal users are small fan-fiction or role-playing communities, such as \sub{/r/HarryPotterRP} or \sub{/r/randomsuperpowers} (a community where users construct individualized superhero identities). 
Loyal communities are also more active, where activity is measured as the average number of comments\footnote{All comments; not only top-level comments.} made per user  ($p<10^{-5}$,  U Test, $4.13\times$ increase). 
However, loyal communities are not necessarily growing. 
If we measure the logarithmic growth rate in subscriber counts\footnote{Subscribers see a subreddit's content on their home page.} for all communities, this value exhibits only a mild positive correlation with loyalty-rates (Spearman's $\rho=0.13, p<10^{-6}$). 

The user-community relation of loyalty is also reflected in some basic properties of user-user interactions.
In particular, the dynamics of conversation threads in loyal communities are noticeably distinct: they tend to be longer ($p<10^{-7}$,  U Test, $7.6\%$ median increase) while containing fewer unique contributors ($p<10^{-5}$,  U Test, $2.5\%$ median decrease).

\subsection{Topics of loyal communities}\label{topical}

\begin{figure}[t!]
\centering
\includegraphics[scale=1.0]{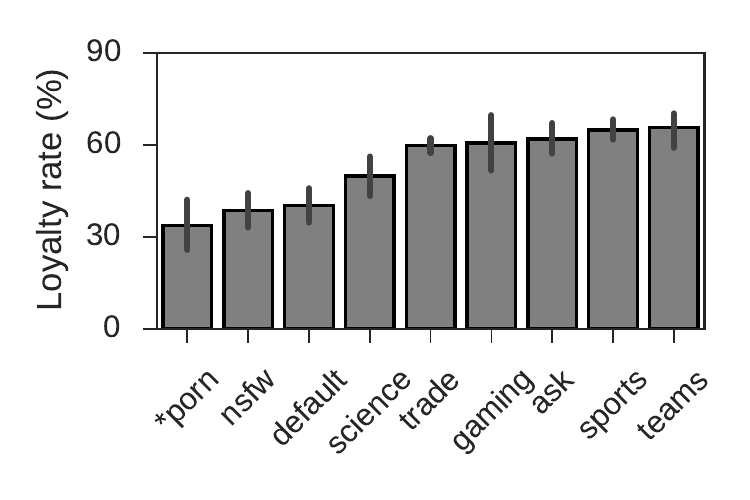}
\vspace{-20pt}
\caption{{\bf Loyalty rates by community category.} Most categories were scraped from /r/ListOfSubreddits, while the category labels containing ``*'' were generated by matching on subreddit names fitting the specified pattern.  Note that ``*porn'' are image-sharing communities like \sub{/r/EarthPorn}, not pornography; the ``nsfw'' category contains pornography and other explicit content. 99\% bootstrapped CIs are shown.}
\label{fig:categories}
\end{figure}

Certain topics, such as sports, tend to engender high rates of loyalty, while other topical categories do not contain many loyal communities. 
Figure \ref{fig:categories} shows the distribution of loyalty-rates by topical category.\footnote{Note that this categorization is far from exhaustive; we obtained topic categories for 20\% of the communities in our dataset.}
Subreddits about sports (e.g., \sub{/r/Cricket}) or specific sports teams (e.g., \sub{/r/Browns}) are by far the most loyal. 
``Default'' subreddits---topically broad communities,  like \sub{/r/pics}, that new users are automatically subscribed to---generally fail to retain loyal users, as do subreddits that are dedicated to sharing images (e.g., \sub{/r/EarthPorn}).

Together these results suggest that loyal communities tend to have specific, focused 
interests, such as a favorite sports team.
In contrast, large topically-diffuse communities, like /r/news or /r/pics, generally fail to retain loyal users. 
The existence of such consistently loyal topics suggests that external identity-based attachment (e.g., to a strong common interest, such as a sports team) may be an important driver of loyalty.

\subsection{Loyalty in interaction networks}\label{sec:networks}
\begin{figure*}[t!]
\centering
\includegraphics[width=1.0\textwidth]{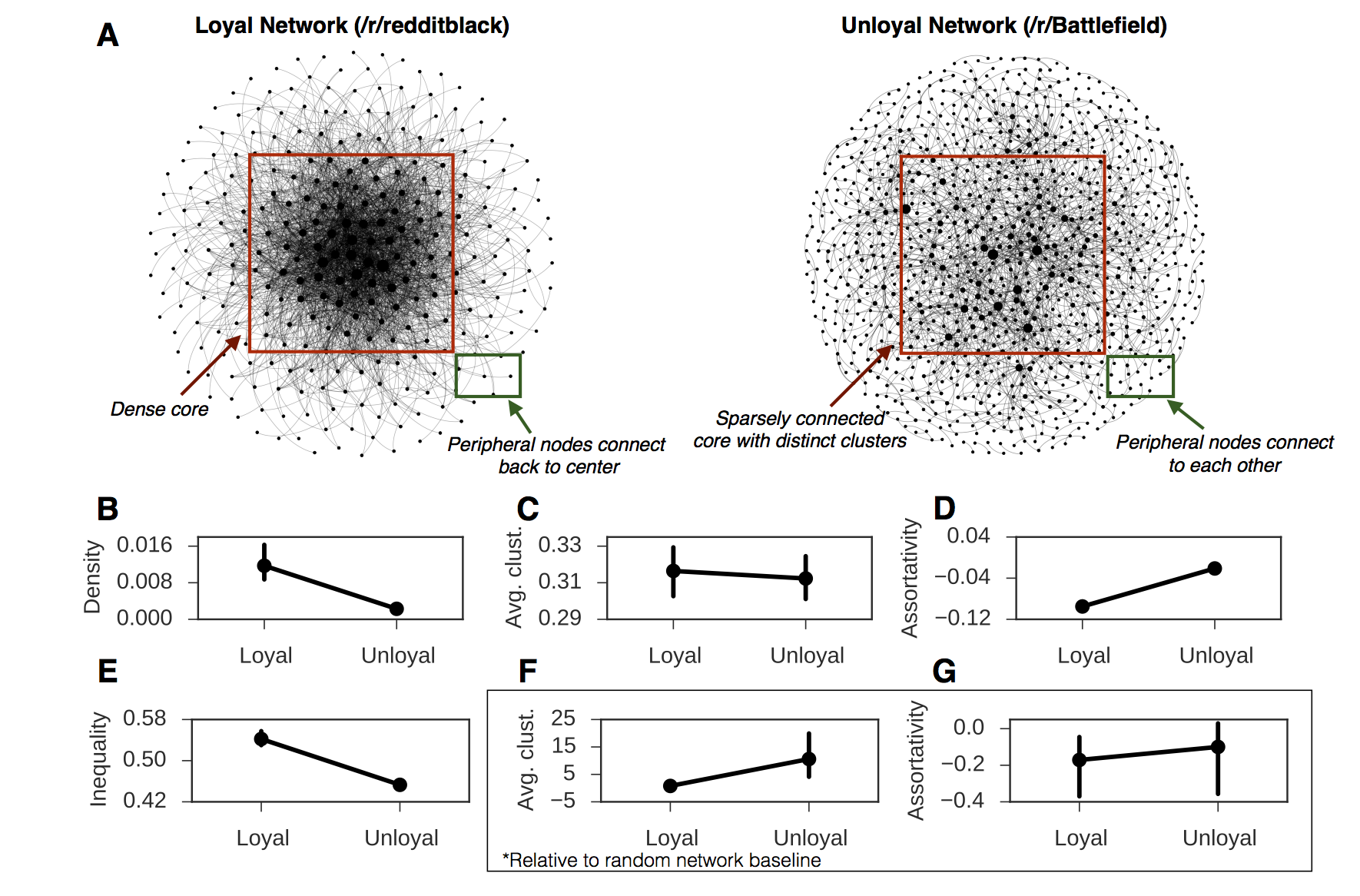}
\vspace{-15pt}
\caption{\textbf{Loyalty reflected in community interaction networks.} Networks of user interactions reveal how loyalty manifests at the collective, community level. \textbf{A}, Example loyal and unloyal networks built from interactions in March, 2014 in two war-related video game communities. \textbf{B-D}, Differences in network statistics for the empirical networks. \textbf{E}, Difference in activity inequality (measured using the Gini coefficient). \textbf{F-G}, Network statistics relative to a null configuration model baseline. Bootstrapped 99.9\% confidence intervals are shown.}
\label{fig:network}
\vspace{-10pt}
\end{figure*}

We find that loyalty is strongly reflected in the patterns of user-user interactions and that these structural markers are predictive of loyalty even after conditioning on a community's topical focus.
In particular, we show that loyal communities have denser interaction networks, less local clustering, and are less assortative by activity level, indicating that loyal communities tend to be more tight-knit and cohesive (i.e., less fragmentation into sub-groups). 

For each community, we construct monthly interaction networks where users are connected if they comment in the same linear comment chain within three comments of each other (i.e., separated by at most two comments).\footnote{Networks and details available at: \url{http://snap.stanford.edu/data/web-RedditNetworks.html}. For computational reasons, we only consider users who made at least 50 comments in 2014 when constructing these networks (roughly the top-20\% of users).
We also replicated key results with direct-reply networks, where users are only connected if one user replied to the other's comment.}
We can reasonably assume that two users who comment in such proximity interacted with each other, or at least directly with the same material.

\xhdr{Analysis of the empirical networks} 
Figure \ref{fig:network}.A-D shows two example networks along with some important statistics highlighting how loyalty is reflected in user-user interactions.
Note that the two communities in Figure \ref{fig:network}.A are both dedicated to particular video games, suggesting that differences in structural markers of loyalty exist between communities with similar topics (a point that we return to below). 

Loyal networks have significantly higher edge density (Fig.\@ \ref{fig:network}.B), even after controlling for activity levels, meaning that the average user in a loyal community interacts with a greater number of other users.
To test differences in density while controlling for activity, we compare matched pairs of communities that are closest in their activity levels (i.e., have similar fractions of comments per user) 
and discarded pairs that differ by more than one-tenth of a standard deviation. 
In this activity-matched setting, we find a significant disparity in edge density between loyal and unloyal communities ($p<10^{-4}$, Wilcoxon Test).

We did not, however, find any significant difference in the average clustering coefficient for the empirical networks (Fig.\@ \ref{fig:network}.C).

The most salient signal in the empirical networks is a difference in activity assortativity (Fig.\@ \ref{fig:network}.D; $p < 10^{-5}$, U test). 
In communities that foster loyalty, highly active users tend to engage with others who have a wide variety of activity levels, while in non-loyal communities users tend to comment near other users of similar activity levels.
Loyal communities also exhibit far higher rates of inequality in their activity levels, measured by the standard Gini coefficient (Fig. \ref{fig:network}.E, $p<10^{-7}$, U test).
Thus, loyal communities have skewed activity distributions,  but they are still \emph{inclusive} in that their highly active ``leaders'' engage with the entire community.

\xhdr{Comparison with a null model} 
The raw differences observed in the empirical networks highlight important ways in which loyalty is reflected in user-user interactions. 
However, these raw contrasts alone do not reveal whether these differences are simply emergent properties of the underlying community structure (e.g., the degree distribution), or whether users in these communities are actually interacting with each other in fundamentally different ways. 
Clustering coefficients, for example, are known to be correlated with node degree in real-world networks \cite{soffer2005network}.
Thus, we compare our networks to suitable null models to control for such confounds.

In particular, we address this issue by comparing the interaction networks to randomly generated networks that have the same degree distribution.
These random networks are generated by randomly rewiring edges while maintaining node degrees, with the number of rewiring iterations set at $10^{4}\times$ the edge count in the empirical network. 
For each community, we compute statistics by taking the relative difference of the median monthly empirical statistic compared to the median monthly null statistic. 

We find that after this control, communities that foster loyalty are significantly less clustered than non-loyal ones (Fig. \ref{fig:network}.F).
Thus, despite the fact that loyal communities have exceptionally dense interactions, they contain fewer closed triads than one would expect. 
Closer inspection reveals that this result is primarily driven by large, sparsely-connected unloyal communities, like \sub{/r/Games} or \sub{/r/aww}. Such communities tend to have extremely high relative clustering, indicating that users in these communities tend to fragment into local clusters.  

Interestingly, we also find that after comparing against a baseline random network, the significant difference in assortativity disappears: 
loyal communities do not have significantly different assortativity levels compared to their random counterparts (Fig. \ref{fig:network}.G, $p=0.15$, U test). 
This implies that users in loyal communities do not actually seek out dissimilar others; rather, they tend to interact with dissimilar others as a consequence of the community's underlying activity and their own activity.

\xhdr{Community structure versus topic}
A key question is how important these structural network features are, relative to community topic, in determining loyalty; it could be that communities that focus on loyal topics, like sports, also just happen to exhibit the structural features described above.  
Since we have topical categorizations for only a small subset of communities (from /r/ListOfSubreddits), we lack the statistical power necessary to directly control for topic. 
However, we can show that for an individual community the structure of its interaction network at particular point of time is predictive of its {\em future} loyalty, even after controlling for the community's current loyalty rate.
If the correlations between community structure and loyalty discussed above were simply the consequence of community topic, this predictive relationship would not exist (since the community's topic, or purpose, is assumed to be stable over time).

We operationalize this idea using a linear mixed model analysis \cite{mcculloch_generalized_2001}, where we regress a community's loyalty rate at time $t+1$ against its current loyalty rate at time $t$ and other relevant community-level features. 
We use a mixed model with random intercepts per community to account for correlated errors and include fixed effects for each month to control for seasonality. 
We run this analysis independently for the two key network features discussed above: network assortativity (raw values) and network clustering (relative to a random model).
This analysis shows that both network features are significant predictors ($p < 10^{-5}$, Holm-Bonferroni corrected Wald's Z-tests).
After controlling for current loyalty rates, a one 
standard deviation decrease in relative clustering is associated with an absolute future loyalty rate increase of $1.6\%$, while an analogous decrease in assortativity is associated with a $2.6\%$ increase.
This means that changes in community structure predict---though not necessarily cause---changes in loyalty, which is strong evidence suggesting that community structure is related to loyalty, independent of topic.

%% file: 040individual.tex
\medskip
\medskip
\section{User-level Loyalty}\label{sec:individuals}

We have shown that certain types of communities tend to foster loyal behavior and that loyal communities exhibit characteristic patterns in their user-user interaction networks.
In this section, we analyze loyalty at the level of individual users. 
We show that loyalty manifests in remarkably consistent ways across a diverse range of communities and that these markers of individual-level loyalty are present even in users' very first contributions to a community. 
 
To reason about our findings in a multi-community setting, 
throughout this section we will say that an effect holds in a particular direction in $X$\% of subreddits, and report the $p$-value under a binomial test where positive and negative outcomes correspond to subreddits where the effect holds in that direction, or the opposite, respectively.

\subsection{Post selection}\label{sec:post_selection}

We study the posts that loyals and vagrants choose to comment on, as a proxy for understanding the {\em tastes} of each user type. 
Across a wide range of communities, we find that loyal users tend to engage with less popular and more esoteric content.

To prevent a few very active commenters from dominating our analyses, we sample posts by randomly selecting $100$ loyals and vagrants per subreddit, then drawing one commented-on post per sampled user, referring to the resultant post collections as {\em loyal-} and {\em vagrant-selected} posts respectively. 

\xhdr{Post popularity} First, we compare the quantity of community feedback given to posts that loyals and vagrants tend to select. We find that in {\em all} subreddits, loyals respond to lower-scoring posts than vagrants; while in 95\% of subreddits ($p < 10^{-9}$), vagrant-selected posts get more comments overall. 
These results suggest that loyals inherently have interests beyond what is currently popular in a community, and may play the important role of surfacing new content that has not yet received community feedback.
One explanation for this effect is that because Reddit displays posts to users ranked by score, loyal users who are more active within a community inevitably navigate further down this ranking and engage with lower-scoring posts. However, we find that users who {\em eventually} become loyal tend to write even their {\em first} comment to a community on less popular posts, compared to vagrants, before any further activity (see also Section \ref{sec:future_classification}).

\xhdr{Content preference}
To explore this post selection process in more depth, we examine the {\em content} of posts that attract loyal and vagrant users.
We show that in addition to commenting on lower-scoring posts, loyal users also tend to prefer posts with more {\em esoteric} content. 

We quantify a post's {\em esotericity} as follows: for a particular noun $\mathcal{N}$, we compute the inverse document frequency (IDF) of $\mathcal{N}$ over all of the posts in that month. The {\em esotericity} of a post $E$ is then the mean IDF of nouns in the post.\footnote{We ignore nouns that only occur once.} 
We focus on nouns to ensure we measure post content as opposed to linguistic style.

Averaging $E$  across posts in our loyal- and vagrant-selected samples, we find that loyals tend to select  more esoteric posts than vagrants in 71\% of subreddits ($p < 10^{-9}$).
Thus loyal users not only select lower-scoring posts, they also tend to comment on content that is more esoteric. 

\begin{figure}
\centering
\includegraphics[width=1.0\columnwidth]{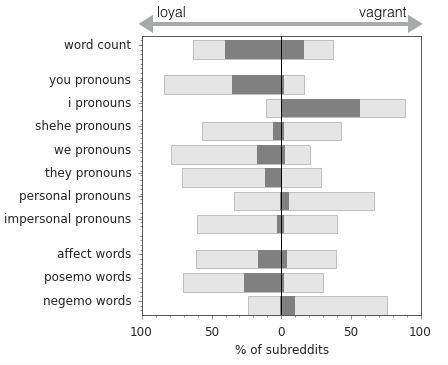}
\caption{\textbf{Linguistic features of comments written by loyal and vagrant users.} Light bars indicate percentages of subreddits where a feature is exhibited more by loyal vs vagrant commenters; dark bars indicate the proportion of subreddits in which this effect is significant (Wilcoxon signed-rank test at the $p < 0.01$ level). \label{fig:comment_feats}}
\end{figure}

\subsection{Linguistic style of comments}\label{sec:user_language}
Having compared how loyals and vagrants select posts to comment on, we now turn to understanding how their comments are written stylistically. To control for the effect of selecting different posts, we construct a dataset of pairs of comments which are responses to the same post, where one comment in the pair is written by a loyal and the other by a vagrant. The median subreddit contains 3065 such pairs. 

We characterize loyal and vagrant comments in terms of linguistic features capturing a comment's lexical style. Motivated by studies of socialization and engagement in online communities \cite{cassell_language_2005,nguyen_language_2011,danescu-niculescu-mizil_no_2013} we explore features relating to comment length (verbosity), distribution of personal pronouns and affect words,
comparing how much each feature is exhibited by the loyal and vagrant comment in each pair.\footnote{We use pronoun and affect word counts from the standard LIWC lexicon \cite{tausczik_psychological_2010}.}
The full set of features is listed in Figure \ref{fig:comment_feats}.

We observe multiple stylistic markers of loyalty in the language of comments which manifest across subreddits, as seen in Figure \ref{fig:comment_feats}; strikingly, these effects emerge even after controlling for the choice of the post responded to. For instance, in 84\% of subreddits, loyals contribute more verbose comments. 

We also note interesting contrasts in the personal pronouns that loyals and vagrants tend to use. After normalizing for length, we find that in 87\% of subreddits, comments authored by vagrants tend to contain more {\em I} pronouns, while loyal users tend to comment with more {\em you} pronouns and more {\em we} pronouns, in 85\% and 79\% of subreddits, respectively. Such preferences echo findings from sociolinguistics and could be attributed to loyal users more strongly identifying with the community \cite{chung2007psychological,sherblom1990organization}.

\subsection{Predicting loyalty from first contributions}\label{sec:future_classification}
\begin{figure}
\centering
\includegraphics[width=1.0\columnwidth]{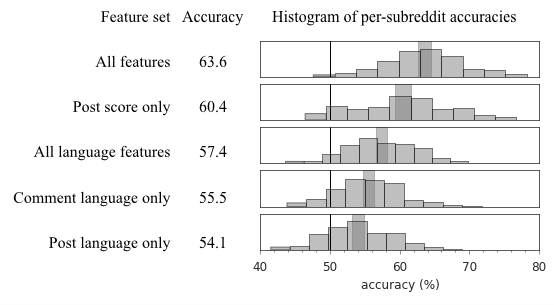}
\caption{\textbf{Predicting loyalty from first contributions.} Histograms depict test accuracies per subreddit, and average accuracy, for predicting future loyalty given a small window of 3 comments at the start of a user's activity in a community. Shaded regions show bootstrapped 95\% confidence intervals around the mean.\label{fig:future-classification}}
\end{figure}

We now consider the prediction task of determining whether a user will become loyal to a community based on their first comments to that community.
The goal of this prediction task is demonstrate that loyalty can be inferred from a user's initial contributions to a community. 
As features we use the score of the posts that the user comments on (Section \ref{sec:post_selection}) as well as the linguistic features defined in Section \ref{sec:user_language}, which we apply to both the text of the user's comments and the posts they reply to. 
 We balance between a positive class of users who become loyal within 2 months of arrival, and users who never become loyal beyond their initial activity. To focus on evaluating the predictive power of these small snapshots, we train one classifier {\em per community} on the initial $k=3$ comments of users who make their first contribution in January to June of 2014, and predict on users arriving to that community in July to October.
 We average feature values over the comments by each user, and enforce that users in both classes must have at least $k$ comments. 
 We use random forest classifiers from the scikit-learn package \cite{pedregosa_scikit_2011} with ensembles of size 100, setting the minimum number of samples required to split a node to 10.

Several groups of features are significantly predictive of future loyalty in many subreddits (Figure \ref{fig:future-classification}). In particular, using all features, 86\% of subreddits have accuracies significantly above the random baseline (averaging 63.6\% accuracy). We note that the relatively strong performance of a classifier that just considers post score suggests that loyal users already seek out unpopular posts early in their lifespan.  We also see that linguistic features can predict better than random in 58\% of subreddits (averaging 57.4\% accuracy). This indicates that in many communities, loyal users have intrinsic affinities for particular elements of linguistic style that manifest very early in their relationship with the community. 
 
 Importantly, all the above markers of loyalty are present in users' first few contributions, meaning that these markers are not simply explained by differences in activity levels. In fact, using these features we can still predict better than random in a majority of communities even when using only the first contribution (k=1).

\subsection{Generalizing across communities}\label{indivdual_classification}
\begin{figure}
\centering
\includegraphics[width=1.0\columnwidth]{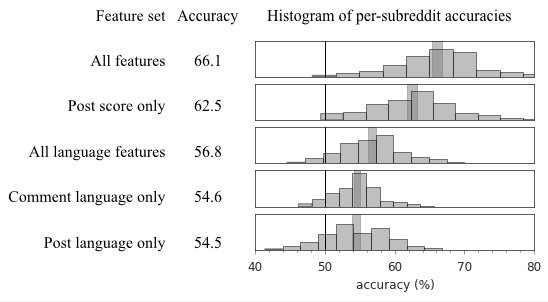}
\caption{
\textbf{Generalizing across communities.}
Histograms depict cross-validation accuracies per subreddit, and average accuracy, in a leave-one-community-out setting classifying loyal and vagrant-authored comments. Shaded regions depict bootstrapped 95\% confidence intervals.\label{fig:user-classification}}
\end{figure}
We have seen a number of features that distinguish loyal users, and that loyalty can be detected from users first contributions to a community.
In this task, we investigate the degree to which these features of loyalty generalize across distinct communities.  

To this end, we consider a {\em leave one community out} prediction setting, 
where the task is determining whether a single comment was written by a vagrant or by a loyal, based on the same features and classifier set-up as in Section \ref{sec:future_classification}.  
In this setting, for each cross-validation fold, we train random forest classifier on all but one community and then predict whether comments in the unseen community were made by loyal or vagrant users.
This setup thus explicitly tests how well the different features of loyalty generalize across communities. 
We construct a balanced dataset by sampling an equal number (250) of loyal and vagrant comments from each of the subreddits considered above. In this section, we do not restrict to first contributions.

We observe that all features generalize well across the vast majority of communities (Figure \ref{fig:user-classification}). Notably, when using all features, 96\% of the subreddits have accuracies significantly above the random baseline ($p <.05$ by a binomial test) with an average accuracy of 66.1\% over all subreddits.  This shows that there are consistent behavioral patterns among loyals, enabling us to infer signals of loyalty over many communities. In addition to these general features, there are also likely to be indicators of loyalty that are highly specific to particular communities. Understanding how these community-specific markers systematically vary across a multi-community space is an interesting avenue for future work \cite{zhang2017typology}.

%% file: 070conclusion.tex

\section{Discussion}

In this work we operationalized loyalty as a user-community relation. 
By doing so, we were able to reveal how loyalty is reflected in the structural properties of user-user interactions and how it manifests in consistent user behaviors across a diverse range of communities.
We exploited this consistency to predict future user loyalty, using only static snapshots of user activity.

\xhdr{Implications for community maintenance}
Our results highlight the important role loyalty plays in community dynamics. 
We found consistent behavioral differences between loyal and vagrant users, in terms of both content they generate and the content they engage with, and revealed that these differences emerge very early in a user's interaction with a community. 
Community maintainers may want to convert vagrant users into loyal ones, but our results suggest that this will require carefully designed entry points; simply optimizing for content that engages loyal users will not convert vagrant users, since they are engaging with fundamentally distinct content. 
More generally, this divide in user interests and engagement patterns suggests that maintainers may need to explicitly balance or optimize the distribution of content that appeals to core, loyal users, compared to content that is attractive to outsiders. 

\xhdr{Limitations and future work}
One important limitation of our work is that we are not privy to individual-level motivations for why users join, and become loyal to, certain communities. In particular, we cannot completely account for the role of extrinsic factors such as reward systems that might contribute towards encouraging ostensibly loyal behavior. For example, we cannot fully disentangle the behavioral markers of loyalty from social dynamics related to the Q\&A-nature of many discussion boards. Even in communities that are not explicitly Q\&A-forums, novice users tend to ask for advice, while long-term users tend to provide it \cite{wang2016learning}; loyal behavior may be positively reinforced by the social capital gained from answering novices’ questions. A promising avenue for future work is to combine large-scale computational analyses with focused surveys of user motivation, in order to more comprehensively understand the psychological motivations behind loyal behavior. 

Further work on a wider variety of platforms is also needed to fully understand the extent to which our results generalize beyond Reddit. Other platforms like Wikia and StackExchange also offer users the opportunity to engage with a variety of communities and possibly establish loyal relationships with some of them. Many such sites also have richer sets of platform-wide affordances such as reward and reputation mechanisms, and further analyses could explore the influence of these various mechanisms in driving and shaping the nature of loyalty.